\documentclass[floatfix,pra,reprint,amsmath]{revtex4-1}
\usepackage{hyperref}
\usepackage[utf8]{inputenc}
\usepackage{graphicx}
\begin{document}

\title{Optimized Bose-Einstein-condensate production in a dipole trap based on a 1070-nm multifrequency laser: Influence of enhanced two-body loss on the evaporation process}
\author{T. Lauber}
\author{J. K\"uber}
\author{O. Wille}
\author{G. Birkl}
\affiliation{Technische Universit\"at Darmstadt, Institut f\"ur Angewandte Physik, Schlossgartenstraße 7, D-64289 Darmstadt, Germany}
\date{\today}
\begin{abstract}
We present an optimized strategy for the production of tightly confined Bose-Einstein condensates (BEC) of $^{87}$Rb in a crossed dipole trap with direct loading from a magneto-optical trap. 
The dipole trap is created with light of a multifrequency fiber laser with a center wavelength of $1070\,$nm. Evaporative cooling is performed by ramping down the laser power only. 
A comparison of the resulting atom number in an almost pure BEC to the initial atom number and the value for the gain in phase space density per atom lost confirm that this straightforward strategy is very efficient.
We observe that the temporal characteristics of evaporation sequence are strongly influenced by power-dependent two-body losses resulting from enhanced optical pumping to the higher-energy hyperfine state. We characterize these losses and compare them to results obtained with a single-frequency laser at $1030\,$nm. 
\end{abstract}
\pacs{67.85.Hj, 37.10.Gh}
%67.85.-d Ultracold gases, trapped gases
%67.85.Hj Bose-Einstein condensates in optical potentials 
%37.10.Gh Atom traps and guides 
\keywords{Bose-Einstein condensation, optical trap, dipole trap, collisional losses, cold collisions}
\maketitle
%***************************************
\section{Introduction}
%***************************************
Since the first experimental realizations of a Bose-Einstein condensate (BEC) in a dilute atomic vapor \cite{Anderson1995,Davis1995}, a variety of different experimental configurations have been developed for efficient BEC production. Besides rf-induced evaporation in magnetic traps, several approaches using optical dipole traps have been implemented starting with the work of \cite{Barrett2001}.
The advantages of condensing atoms in optical rather than magnetic traps are the relatively simple setup, the potential to directly transfer the BEC into optical trapping or guiding configurations \cite{Bongs2001}, and  the possibility to simultaneously investigate atoms in different spin states \cite{Stamper-Kurn1998b,Barrett2001} or in states without magnetic moment \cite{Kraft2009}.
A disadvantage lies in the fact that here the standard method to selectively remove the atoms with the highest energy during evaporative cooling is to reduce the trap depth \cite{Adams1995} by decreasing the laser power. This, in general, reduces the trapping frequencies as well. As a consequence, efficient rethermalization may not be possible at the end of the evaporation if very low laser power is required.

Many groups generating BECs in dipole traps use CO$_2$-lasers \cite{Barrett2001,Cennini2003,Kobayashi2006,Gericke2007} in single-beam or crossed-beam configurations. Slight disadvantages result from the high power required for the laser wavelength of $10.6\,\mu$m and from the need to implement optical materials transparent at this wavelength which might complicate the optical setup. 
An alternative consists in the use of a laser with a wavelength close to $1\,\mu$m which is also covered by commercially available high-power laser systems. Advantages of this wavelength are the reduction of the required laser power and the possibility of using the optical components of the dipole trap for laser cooling and manipulation of the atoms before and after the BEC production phase as well. 
On the other hand, complications have been encountered in experiments using near $1\,\mu$m lasers for production of a BEC of $^{87}$Rb atoms: unexpected high atom losses have been observed especially when using cost-efficient high-power multifrequency laser systems, such as fiber lasers, at this wavelength. 
As a consequence, only a limited number of experiments using lasers at $1\,\mu$m wavelength for the  confining potential for Bose-Einstein condensation of $^{87}$Rb have been reported \cite{Kinoshita2005,Hung2008,Couvert2008,Clement2009,Arnold2011,Zaiser2011}, and frequently complex experimental strategies have to be implemented. For example, shiftable lenses \cite{Kinoshita2005} allow a compression of the trap during evaporation to compensate for the reduced optical power. Dimple traps consisting of two beams, a single beam with a large volume and a second more tightly focused beam which is superimposed during a later stage of evaporation allow a recompression of the atoms \cite{Clement2009,Kraemer2004}. These methods are based on the possibility to change the phase space density by changing the shape of the potential \cite{Stamper-Kurn1998a,Pinkse1997}.

In this paper, we present an optimized BEC production strategy which waives these additional complexities. Our setup consists of a crossed dipole trap created with the light of a 1070\,nm multifrequency fiber laser. The dipole trap is directly loaded from a standard magneto-optical trap (MOT). Only the dipole laser intensity is changed during evaporation. We present the optimized evaporation path, discuss the strategy for finding it, and show that our configuration produces BECs with high efficiency.
Special attention is given to the effects of the multifrequency spectral distribution of the dipole laser: we present a detailed investigation of the occurring atom losses which confirms the validity of the evaporation strategy. We compare these results to the ones obtained with a single-mode laser of similar wavelength and beam parameters which allows us to verify that not the laser wavelength but rather the spectral characteristics of the laser light are essential for understanding the extra difficulties encountered in some experiments for Bose-Einstein condensation of $^{87}$Rb in $1\,\mu$m wavelength dipole trap configurations.

In the following sections, we first present our experimental setup (section \ref{secsetup}), describe the optimized sequence used for atom preparation and evaporation to BEC (section \ref{secevapseq}), and give evidence for the achievement of BEC in our setup. Section \ref{secloss} is dedicated to the investigation of the occurring loss mechanisms and the resulting consequences for the optimized shape of the intensity ramp during evaporation.

%***************************************
\section{Setup}\label{secsetup}
%***************************************
To create a BEC, we implement a three-stage cooling process: we first decelerate an atomic beam, trap and cool the atoms in a MOT, and finally transfer the atoms into the crossed dipole trap to cool them evaporatively.

Rubidium is heated in an oven and directed as atomic beam by a nozzle and a differential pumping stage into the main vacuum chamber having a pressure of $3.5\times10^{-11}\,$mbar. The atomic beam is decelerated via frequency-chirped laser beams and trapped in a MOT where we accumulate around $4\times10^7$ rubidium atoms in 8\,s loading time.

\begin{figure}
\includegraphics[width=0.95\columnwidth]{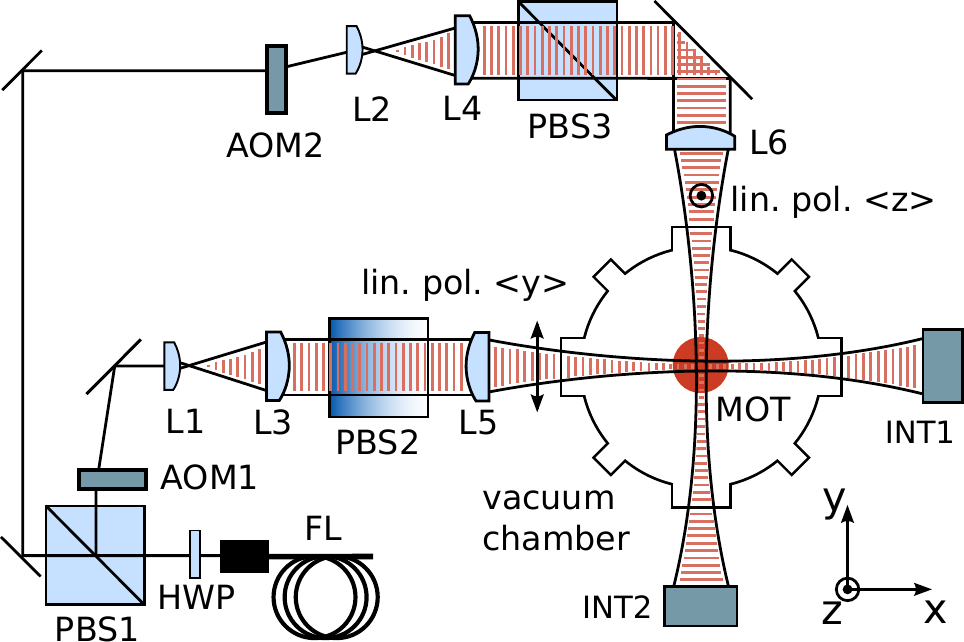}
\caption{\label{figschematic}(Color online) Schematic top view of the dipole trap laser setup. The beam from the fiber laser (FL) is split in two parts at beam splitter PBS1. In each beam line an acousto-optic modulator (AOM1 and AOM2)  controls the intensity. The initial foci are prepared with the lenses L1 and L2 and relayed into the vacuum chamber with achromatic lens pairs (L3,L5 and L4,L6). The polarization of the two beams is fixed with two perpendicularly oriented polarizing beam splitter cubes (PBS2 and PBS3). Monitoring of the beam intensities behind the chamber is achieved by photo diodes INT1 and INT2}
\end{figure}

The dipole trap is created by crossing two focused laser beams perpendicularly (see Figure \ref{figschematic}). The beams are generated by using a linear polarized $1.07\,\mu$m multifrequency fiber laser (IPG YLR-50-1070-LP) with a typical output power of $50\,$W and a spectral width (FWHM) of 2\,nm. We use only $20\,$W for BEC production because higher power caused problems with thermal effects in the optical elements along the path of the light and the remaining laser light is used for creating micro-optical potentials in further experiments, such as discussed in \cite{Dumke2002,Kreutzmann2004,Eckert2006}. The beam is split in two beams with selectable power ratio via the wave plate HWP and the polarizing beam splitter PBS1. An acousto-optical modulator (AOM) regulates the power in each beam line. 
Special care has been taken to avoid interference effects in the dipole trap crossing: we use perpendicular linear polarizations for the beams and assure this by cleaning the polarization with orthogonally oriented beam splitter cubes (PBS2/PBS3) behind the AOMs. This is due to our observation that the AOMs occasionally modify the polarization state when the radio frequency power is changed. To further reduce potentially remaining interference effects, we operate the two AOMs in opposite diffraction orders (first and minus first order, respectively). This results in a frequency difference of $220\,$MHz between the two beams, and no temporal modulation of the total dipole potential on a time scale experienced by the atoms should occur.

In each beam line a laser focus is generated outside the vacuum chamber using lenses L1 or L2, respectively, which is relayed into the vacuum chamber onto the position of the MOT with scale 1:1 by two $f=500\,$mm (L3,L5 and L4,L6) achromatic lenses. This gives the flexibility to change the waist of each beam by changing L1 and/or L2 without the need for further modifications of the setup. 
In order to get the most reliable values for the waists inside the vacuum chamber we measured the trap frequencies by parametric heating \cite{Jauregui2001} together with the laser power. 
The beam waists for all results presented here are $w_1=(41\pm2)\,\mu$m and $w_2=(46\pm2)\,\mu$m respectively. We found this choice to be very effective, presenting a good compromise between high trapping frequencies (for efficient evaporation dynamics) and a reasonably large trap volume (for a large initial atom number). 
With these waists and a total maximum power \footnote{the relative uncertainties in laser power, peak intensity, and trap depth are about 4\% throughout this paper} of $11.7\,$W at the position of the atoms, we obtain a peak intensity of $396\,\mathrm{W/cm}^2$ which corresponds to a maximum total trap depth of $k_B\times595\,\mu$K, and an average trapping frequency $\bar{\nu}=\sqrt[3]{\nu_x\nu_y\nu_z}=1.4\,$kHz \footnote{the relative uncertainties in the trapping frequencies are about 10\%}. The power ratio between the two beams is chosen in a way that both single-beam traps provide the same amount to the total trap depth.
After passing through the experimental chamber, each beam intensity is monitored by a logarithmically amplified photodiode (INT1 and INT2), since it is necessary to accurately measure the power over 3 orders of magnitude during evaporative cooling. In order to prevent saturation of the photodiodes we attenuate each beam by using a lens with a short negative focal length and a beam block with a hole in the center to pass only the small central part of each beam. The signals of the amplified photodiodes are used to actively stabilize the light power in each beam to a computer-controlled value via the AOMs.

An efficient transfer of rubidium atoms from the MOT to the dipole trap is achieved as follows. Already during the MOT loading phase, the dipole trap laser beams are switched on at the maximum power ($11.7\,$W). After the MOT phase, the MOT magnetic field is switched off and loading of the dipole trap is optimized by lowering the intensity of the MOT cooling and repumping light and increasing the detuning of the cooling light \cite{Kuppens2000}. This results in an optical molasses of 90\,ms duration with reduced temperature. At the end of the molasses phase we switch off the repumping light 2\,ms before the cooling light to accumulate the atoms in the lower hyperfine level $(F=1)$ of the ground state $5^2S_{1/2}$ of $^{87}$Rb. In this state the lifetime in the dipole trap is larger, because hyperfine state changing collisions are suppressed \cite{Weiner1999,Grimm1999}. A preparation to a single magnetic sublevel with $\left|m_F\right|=1$ is available but was not applied in the measurements presented here. With this loading method we typically obtain $3.5\times10^5$ atoms in crossed-beam section of the dipole trap. Atoms initially loaded into the single beam 'wings' of the trap are omitted as they are rapidly spilled during evaporation.
The peak density is $1.1\times10^{13}\,\mathrm{cm}^{-3}$ and the temperature $100\,\mu K$. This gives an initial phase space density of $2\times10^{-5}$ assuming equal atom distribution within the three $(F=1)$ spin states. 
Since we observe extremely high atom losses under these conditions, we reduce the power from $11.7\,$W to $8.8\,$W within five milliseconds and take the resulting trap as starting point for our sequence of
forced evaporative cooling through lowering the intensity of the trap laser beams.

%***************************************
\section{Evaporation Sequence and Evidence for BEC}\label{secevapseq}
%***************************************
At $8.8\,$W of total power, the trap center experiences a peak intensity of $296\,\mathrm{kW/cm}^2$ and a spontaneous scattering rate of $1.5\,\mathrm{s}^{-1}$. The total trap depth is $k_B\times 444\,\mu$K and the average trap frequency is $\bar{\nu}=1.2\,$kHz. Here, we still observe a $1/e$-lifetime of the trapped atoms below one second, which at first sight seems to prevent efficient evaporative cooling. Nevertheless, we could achieve efficient Bose-Einstein condensation under these apparently adverse starting conditions: we observed that the atom losses decrease rapidly when the laser power is reduced and therefore implemented an optimized evaporation sequence.
\begin{figure}
  \includegraphics[width=0.95\columnwidth]{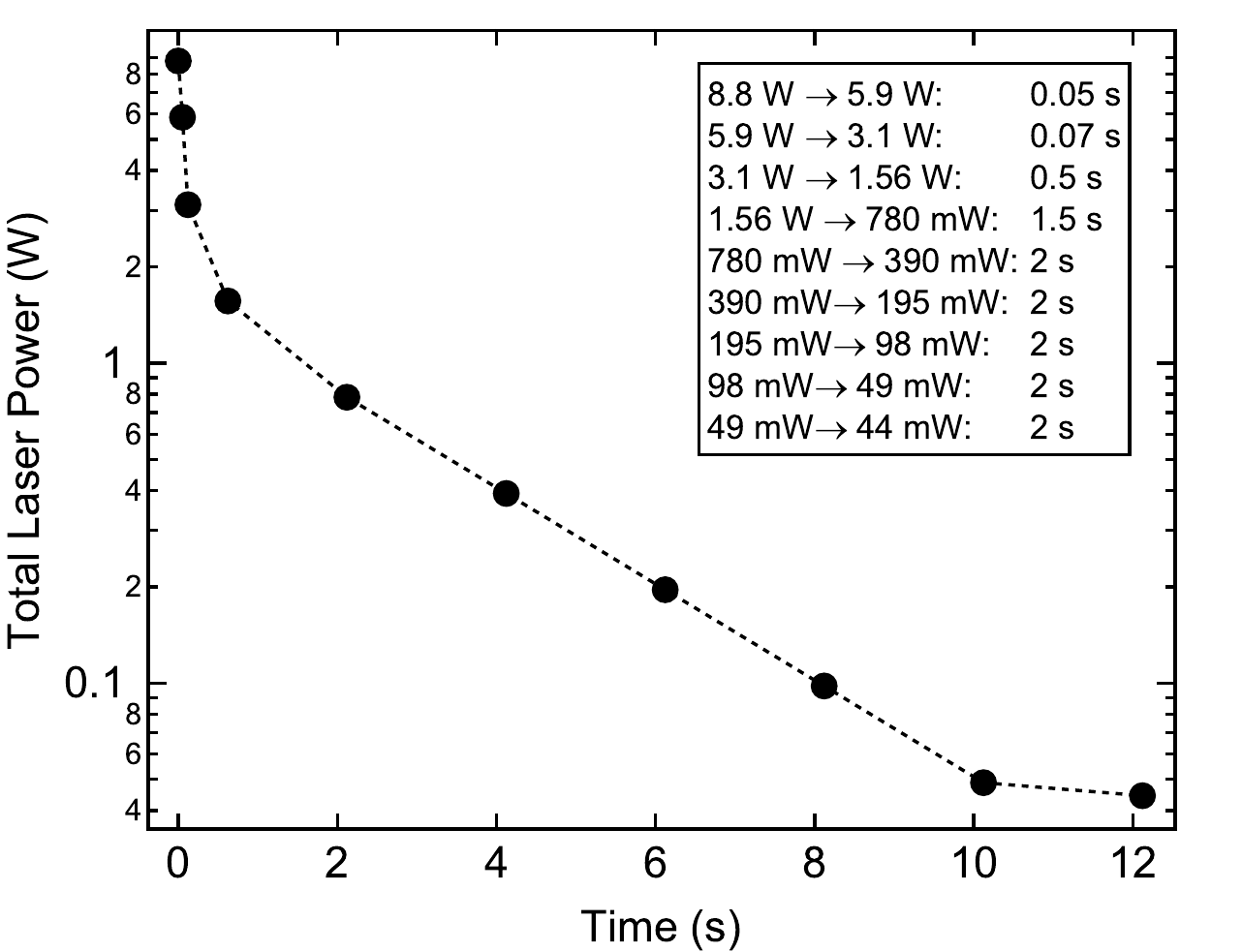}
\caption{Time sequence of the total laser power during evaporative cooling. The inset gives the start and end values of the laser power together with the duration for each linear evaporation ramp.}
\label{figramp}
\end{figure}
Usually, for evaporative cooling in optical dipole traps, a temporal variation of the laser power according to the scaling laws of \cite{OHara2001} is applied. We approximate the resulting function by a series of linear ramps. Each ramp reduces the intensity by a factor of 2. We optimize the duration of each ramp experimentally for highest gain in phase space density. The resulting time sequence of the laser power $P_e(t)$ for our configuration is shown in Fig. \ref{figramp}. We observe that the time constants $\tau_{ramp}$ are much smaller during the first two ramps than in the following segments. This is a consequence of the strong losses occurring at high laser power. During the next two segments, the time constants approach the scaling law behavior which is followed for the subsequent four segments.
The very last evaporation ramp has a smaller slope than the ones before because the power at which we achieve Bose-Einstein condensation is close to the limit where the trap is not able to support the atoms against gravity. This results in a distortion of the trapping potential in the vertical direction which decreases the trapping frequencies and the elastic scattering rate. This increases the time required for rethermalization and makes it necessary to reduce the slope in the final evaporation ramp. 
We end up at the BEC phase transition after an evaporation time of $12.1\,$s, which together with the loading time of the MOT of $8\,$s gives a total cycle time of approximately $20\,$s.

To avoid the high initial losses one might consider to load the dipole trap at a power which corresponds to the point of turnover to the scaling law behavior which corresponds to a power of about $3.1\,$W in the sequence of Fig. \ref{figramp}. 
We tested this by loading the dipole trap at a constant total power of $3.5\,$W and compared the parameters of the resulting atom sample to our original loading scheme followed by the first part of the evaporation sequence ending at a laser power of of $3.5\,$W. We obtained comparable temperatures here as well as after one additional evaporation ramp but for both times the number of atoms in the low-power loading case was only about $40\%$ of the atom number when following the original sequence. This proofs that it is advantageous to use the higher initial laser power even in the presence of high initial losses when making use of an optimized evaporation sequence.

\begin{figure*}
\resizebox{0.92\textwidth}{!}{
\includegraphics{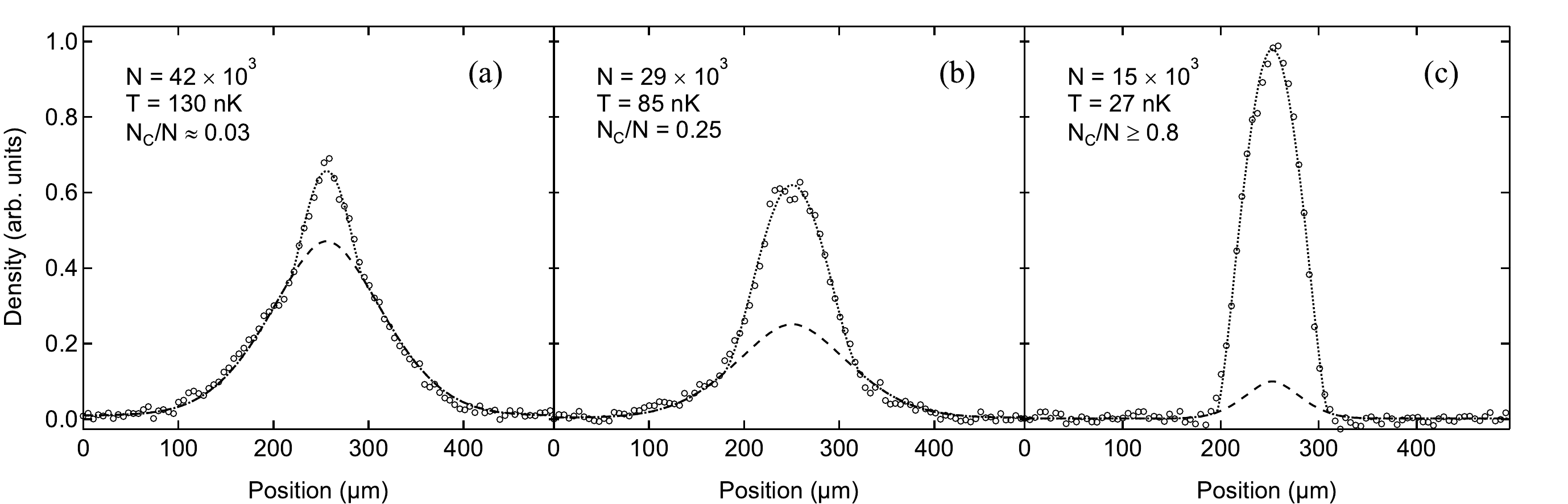}
}
\caption{Cross sections of the atom distribution at different final trap depths after the transition to BEC. (a) small BEC and mainly thermal cloud right below the phase transition, (b) partially condensed cloud, and (c) almost pure BEC with a condensate fraction above 80\%. All data are taken after the same time-of-flight of $20\,$ms. The dotted lines represent bimodal fits to the density distribution while the dashed lines indicate the thermal fraction.}
\label{figbimodal}
\end{figure*}

At the end of the evaporation where we use a total power below 45\,mW (trap depth $k_B \times 2.2\,\mu$K) for the dipole trap, we achieve Bose-Einstein condensation. The critical temperature is around 140\,nK. At the phase transition we are typically left with $4\times10^4$ atoms. A measurement of the resulting bimodal distribution, as a proof for condensation is shown in Fig. \ref{figbimodal}.
If we evaporate to lowest achieved temperatures (Fig. \ref{figbimodal} (c)) we can create almost pure condensates (condensate fractions higher than 80\%) with a temperature below 30\,nK and a total atom number of around $1.5\times10^4$. There, the remaining thermal atom number is hard to determine precisely due to its small fraction and density in the thermal wings of the distribution.

In spite of the large initial losses, this straightforward evaporation procedure is very efficient. We determined two figures-of-merit to confirm this: 
the ratio of the atom number at the start of evaporation to the atom number in an almost pure BEC $N(t=0)/N_c=35$ lies close to the high efficient end of values found in the literature (30 to 100) for evaporation in dipole traps. In addition, the evaporation efficiency
\begin{align}
\gamma_\mathrm{ev}=-\frac{\ln(\rho_{PSD,c}/\rho_{PSD}(t=0))}{\ln(N_c/N(t=0))}\quad,
\end{align}
which compares the gain in phase space density $\rho_{PSD}$ to the reduction in atom number, is $\gamma_\mathrm{ev}\simeq3.5$ which is comparable to values found in the literature ($\gamma_\mathrm{ev}=3.4$ \cite{Hung2008}, $\gamma_\mathrm{ev}=2.8(5)$ \cite{Clement2009}) for highly efficient, but more sophisticated dipole trap evaporation setups.

%***************************************
\section{Analysis of Atom Losses and their Effect on Evaporation}\label{secloss}
%***************************************
\begin{figure}
  \includegraphics[width=0.95\columnwidth]{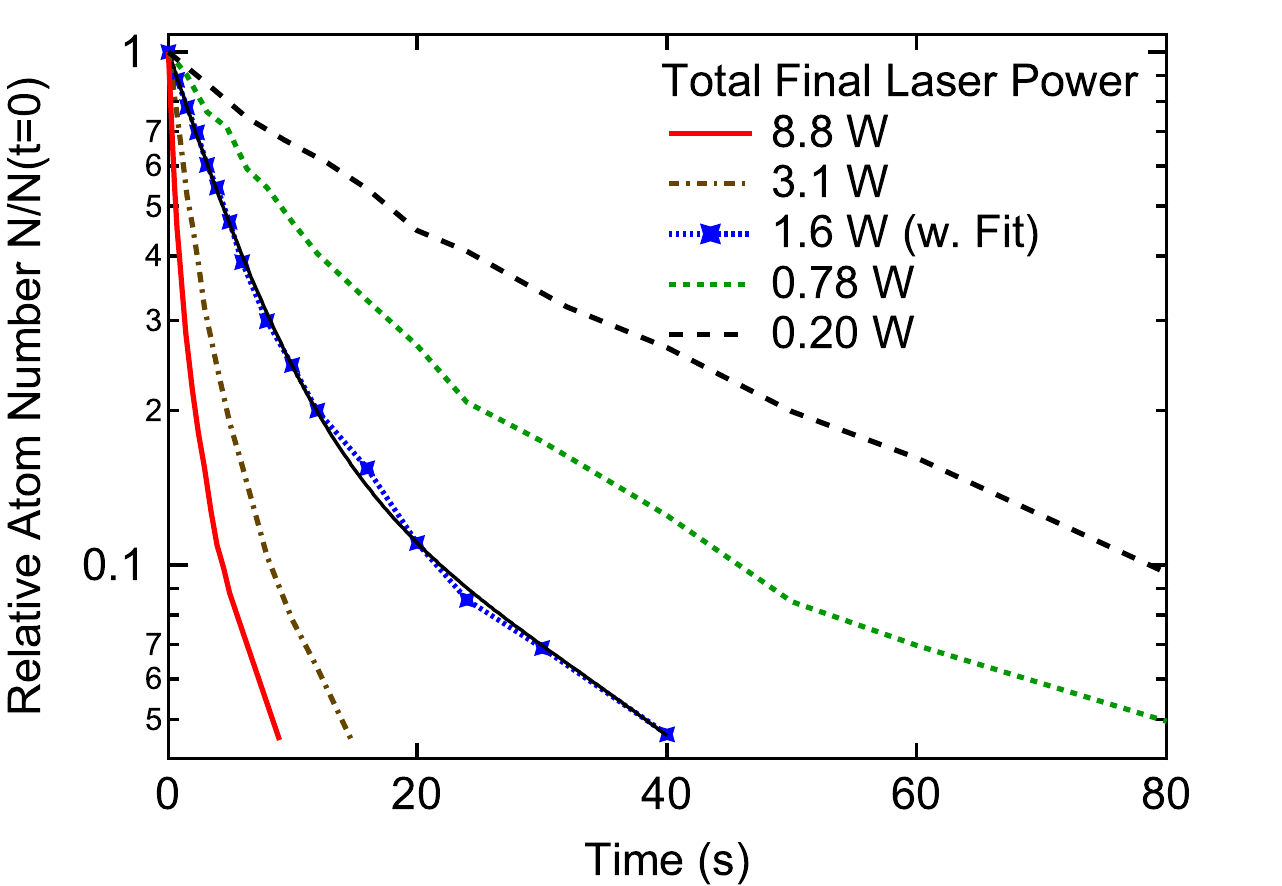}
\caption{(Color online) Atom number decay in the crossed dipole trap for different total final laser powers. 
The decay can be well fitted by a sum of two exponentials (see text) as plotted for the case of $1.6\,$W of laser power where the data (crosses and short dashed curve) are compared to the fit (black line).
The decay curves for a total power below $0.2\,$W show the same behavior as the one for $0.2\,$W and are not plotted.}
\label{figdecay}
\end{figure}

We performed a detailed analysis of the atom losses during different stages of evaporation in order to gain insight into the specific loss characteristics in dipole traps generated by a multifrequency laser and to confirm the experimentally obtained evaporation sequence $P_e(t)$ (Fig. \ref{figramp}). For this purpose, we analyzed the temporal evolution of the atom number in the crossed dipole trap at different laser powers (Fig. \ref{figdecay}). Each decay curve is measured following the evaporation sequence shown in Fig. \ref{figramp} terminated at the indicated laser power which is then kept constant for the decay curve measurement. As a consequence, temperature, atom number, and density at $t=0$ are the same as they would be during our experimental evaporation sequence.
Fig. \ref{figdecay} shows high atom losses at high laser power but also a strong reduction of the loss for decreasing laser power. 
The increase in atom trapping time of more than one order of magnitude is essential for the successful evaporation strategy.

Another obvious feature in the semi-logarithmic presentation is the curvature of the decay curves, especially at high laser power. This is caused by density-dependent many-body losses occurring in addition to single-atom losses. Even for the highest densities, achieved directly after loading, we calculate an atom loss rate due to three-body recombination \cite{Burt1997} only of around 500 atoms $\mathrm{s}^{-1}$. Thus, three-body processes play no significant role in the observed atom loss, and we focus on one- and two-body losses in the further discussion.

\begin{figure}
  \includegraphics[width=0.95\columnwidth]{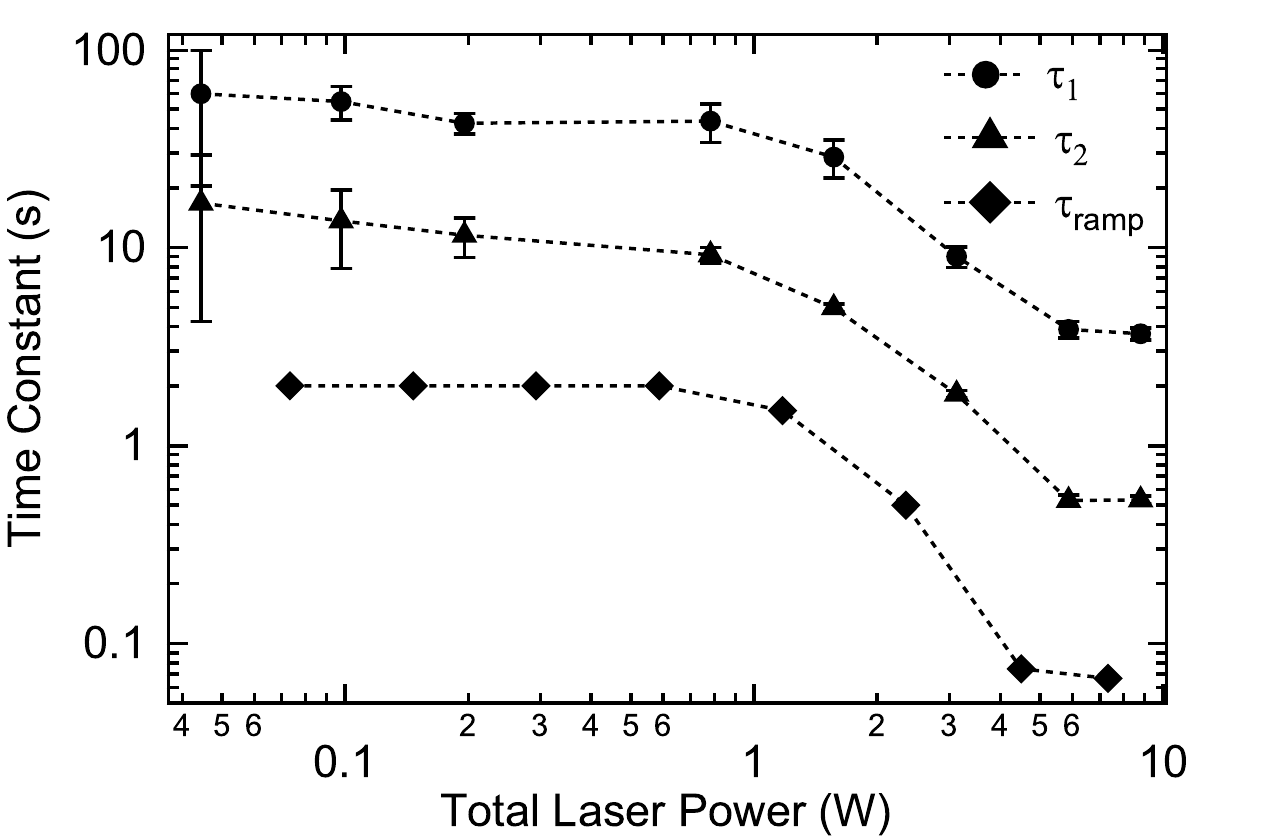}
\caption{Time constants $\tau_1$ and $\tau_2$ for atom loss and the time constant $\tau_\mathrm{ramp}$ representing the duration found experimentally for each ramp reducing the laser power by a factor of 2 as a function of laser power. For the ramp constant, the corresponding laser power is given as the average power during the respective linear ramp.}
\label{figtau}
\end{figure}

Since, here, we are only interested in comparing the time constants for atom loss to the time constants of the evaporation ramps rather than extracting quantitative two-body loss coefficients, we determine the atom loss time constants in Fig. \ref{figdecay} by fitting a sum of two exponential decay curves to the data. 
The extracted time constants are plotted in Fig. \ref{figtau}. 
The smaller time constant, $\tau_2$, is dominated by two-body losses, the longer, $\tau_1$, by single-atom losses due to background gas collisions and heating through photon scattering from the dipole trap laser.

Figure \ref{figtau} also shows the time constant $\tau_\mathrm{ramp}$ which is the time needed to reduce the laser power by a factor of 2 in each linear ramp segment of Fig. \ref{figramp} as a function of the average power during the segment. In our evaporation sequence, no linear ramp segment is long enough to show a significant influence of a decay with time constant $\tau_1$, and atom loss during evaporation is fully dominated by the fast exponential decay with time constant $\tau_2$. Obviously, the ramp durations obtained by the the experimental optimization procedure are proportional to the observed loss constants. The proportionality factor $\tau_2/\tau_\mathrm{ramp}\geq6$ for all ramp segments. The observation of this constant ratio gives another simple and straightforward strategy for choosing the appropriate time constants for evaporation: one can measure the loss constant of the atom ensemble and select the ramp time constant proportional to it. The specific value for the ratio $\tau_2/\tau_\mathrm{ramp}$ we obtained here is only valid for our set of parameters, but following this strategy reduces the requirements on the optimization procedure to optimizing only one single free parameter.

To gain additional insight into the reason for the large atom losses at high laser power, we compared the results given in Fig. \ref{figtau} with an atom loss measurement carried out with a single-frequency laser (ELS Versadisk Yb:YAG, 1030\,nm) in the same setup at a power of $9.5\,$W and comparable values for trap size, atom number, and background gas pressure. With the single-frequency laser, we observed time constants $\tau_2=6.5\,$s and $\tau_1=23.6\,$s. Both time constants are about one order of magnitude larger than the ones obtained for the multifrequency fiber laser and the optimized evaporation sequence does not require the fast ramping at the beginning. Due to these observations we infer that the high initial losses for the fiber laser are not due to the absolute wavelength around $1\,\mu$m, but rather due to the laser's broad frequency spectrum. The different longitudinal modes are distributed over a range of more than 500 GHz with a separation of approximately $15\,$MHz while the line width of the atomic transition in rubidium is $6\,$MHz. It is very likely that each mode of the laser field has a counterpart with a frequency offset close to the hyperfine splitting of the ground state of $^{87}$Rb of $6.834\,$GHz. In this case, two laser photons could drive the transition from the $F=1$ to the $F=2$ hyperfine ground state which has an internal energy higher than the trap depth. In hyperfine state changing collision back to the $F=1$ state, internal energy is converted to kinetic energy. This leads to loss of the colliding atoms \cite{Weiner1999}.\\
\begin{figure}
  \includegraphics[width=0.95\columnwidth]{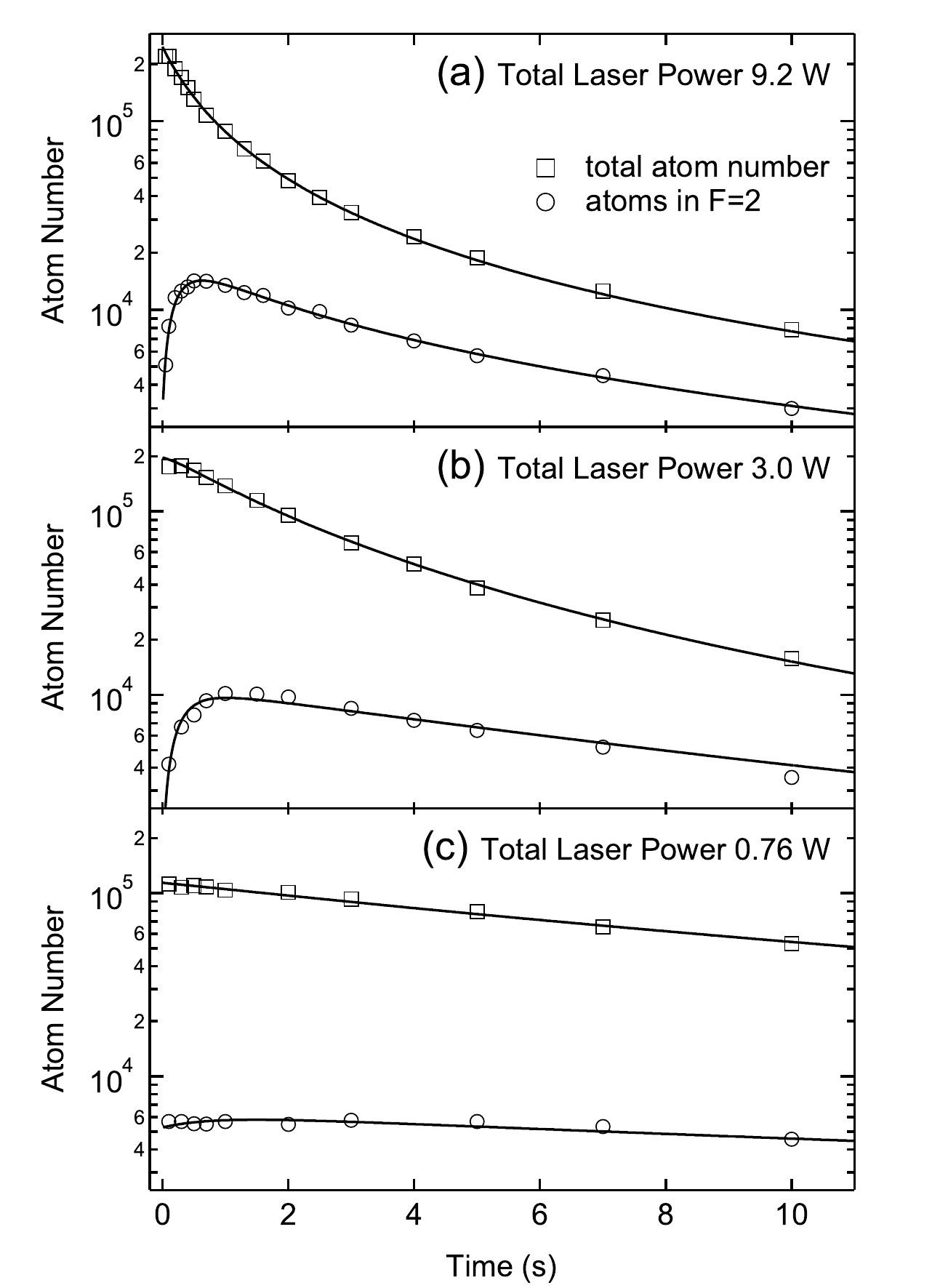}
\caption{Temporal evolution of the total atom number and the number in state $F=2$ together with the resulting fit of the solution of the rate equations (\ref{eqrate}) for a total laser power of (a) $9.2\,$W, (b) $3.0\,$W, and (c) $0.76\,$W}
\label{figf2pop}
\end{figure}

To support this model, we experimentally investigated the dipole trapping light induced transfer of atoms to state $F=2$ and the subsequent hyperfine relaxation by measuring the population $N_2$ in the higher hyperfine level $F=2$ evolving in time. State-selective detection is achieved by absorption imaging on the cycling transition $F=2\rightarrow F'=3$ without additional repumping light. Only atoms in state $F=2$ are able to absorb light. The reduction of detection efficiency due to the missing repumping light was experimentally confirmed to be negligible during the time of the detection pulse. Repeating this measurement with repumping light gives the temporal evolution of the total number of trapped atoms. Results for three different laser powers are presented in Fig. \ref{figf2pop}. For laser powers of $9.2\,$W and $3.0\,$W, we observe a fast initial increase of $N_2$ followed by a decay corresponding to the decay of the total atom number. No additional pumping to $F=2$ is observed for a laser power of $0.76\,$W and below. As can be seen at short times for $9.2\,$W but especially for $3.0\,$W, the fast two-body decay of the total atom number has to be initiated by a transfer of atom population to $F=2$ first. This can be expected due to the significantly larger rates for two-body collisions involving at least one atom in the upper hyperfine state $F=2$.

\begin{figure}
  \includegraphics[width=0.95\columnwidth]{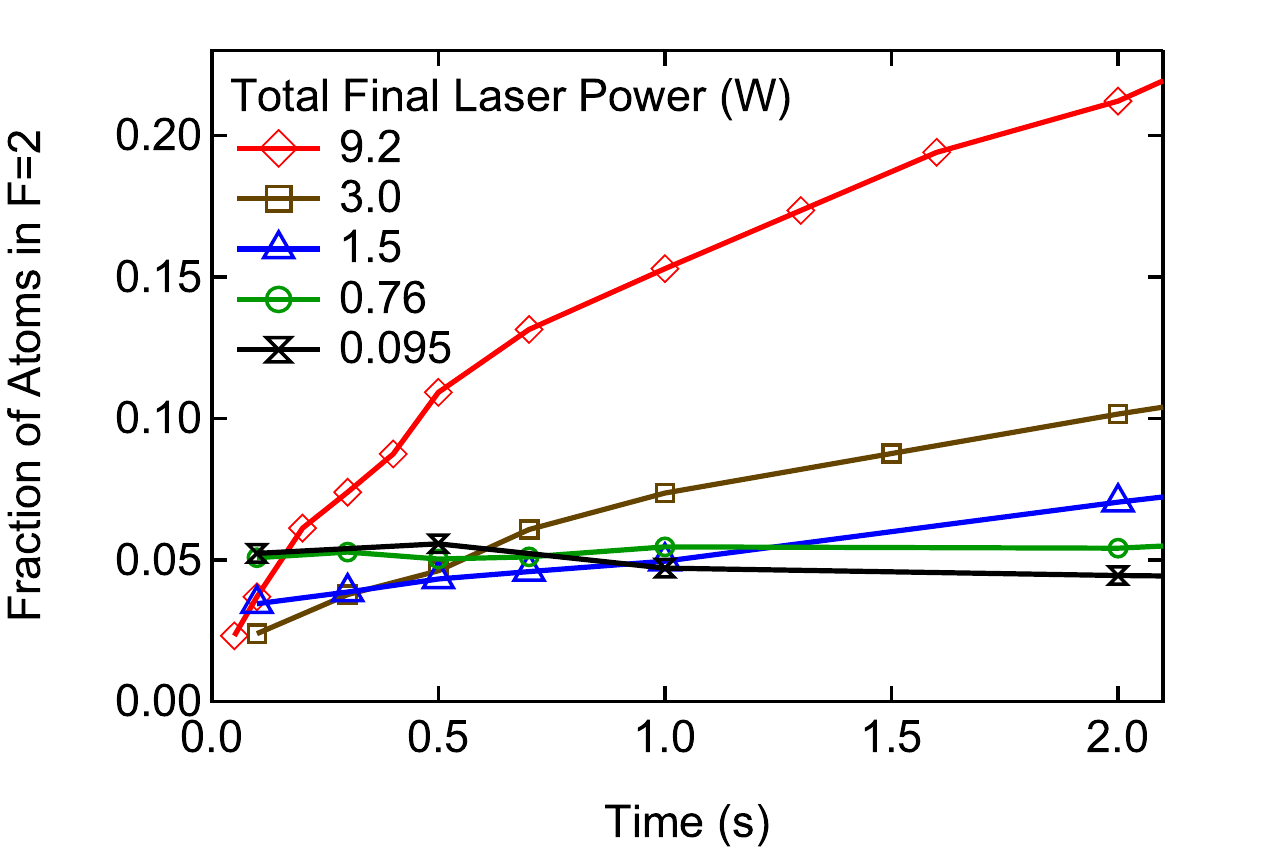}
\caption{(Color Online) Temporal evolution of the fraction of atoms in state $F=2$ for different total final laser powers}
\label{figf2frac}
\end{figure}

A detailed presentation of the temporal evolution of the fraction of atoms in $F=2$ during the first two seconds is given in Fig. \ref{figf2frac}. The non-vanishing initial fraction of atoms occurring in $F=2$ in each curve is not due to imperfect preparation to $F=1$ but rather caused by pumping of atoms to $F=2$ during the evaporation stages preceding the start of each of the presented measurements. Clearly, the fraction of atoms populating the state $F=2$ rapidly increases at high laser power. At $9.2\,$W the fraction rises within two seconds to a value of $21\%$, while at lower power the fraction increases slower and saturates at lower values. The high fraction of atoms in state $F=2$ then induces enhanced two-body loss for  $F=2$ atoms either colliding with each other or with the remaining $F=1$ atoms. This enhanced loss would not occur if all atoms keep their initial state  $F=1$. For a laser power of $0.76\,$W and below, no increase in the fraction of atoms pumped to $F=2$ is observed.

Spontaneous pumping to state $F=2$ is possible by absorption of a photon from the dipole trap laser and emission of a photon with a lower energy. For large detuning, these spontaneous Raman processes are suppressed by quantum interference because the scattering amplitudes via the $D_1$- and $D_2$-lines of rubidium almost cancel \cite{Cline1994}. This effect yields a suppression, compared to the total spontaneous scattering, by a factor of $1000$ with linearly polarized light for transitions from ground state hyperfine level $F=1$ to $F=2$ at a trapping laser wavelength of $1070\,$nm. We calculate the spontaneous Raman scattering rate to be only around $1.5\times10^{-3}\,\mathrm{s}^{-1}$ at $9.2\,$W. Thus, spontaneous Raman scattering cannot explain the observed rapid increase in $F=2$ population. Together with the observation that a single-frequency laser field does not induce large atom losses, we infer that driven two-photon transitions induced by two different modes of the multifrequency laser field are responsible for the significant increase of the atom population in state $F=2$ followed by the higher atom losses due to the larger loss rate for collisions involving $F=2$ atoms.

To describe the evolution of the atom populations in $F=1$ ($N_1$) and $F=2$ ($N_2$), we use a simple rate equation model:
\begin{align}
\begin{split}
\dot{N}_1=&-p(N_1-N_2)-\beta_{11}\frac{N^2_1}{V_\mathrm{eff}}-\beta_{12}\frac{N_1N_2}{V_\mathrm{eff}}\\
\dot{N}_2=&+p(N_1-N_2)-\beta_{22}\frac{N^2_2}{V_\mathrm{eff}}-\beta_{12}\frac{N_1N_2}{V_\mathrm{eff}}\quad,
\end{split}
\label{eqrate}
\end{align}
with $V_\mathrm{eff}=(4\pi k_B T/m)^{3/2}/(2\pi\bar{\nu})^3$ being the effective volume for the calculation of the average density, where $\bar{\nu}$ is the average trapping frequency. We fit the solutions of the rate equations for $N_1(t)$ and $N_2(t)$ to data as in part presented in Fig. \ref{figf2pop} to obtain the relevant parameters, which are the two-body loss coefficients for collisions between two $F=1$ atoms $\beta_{11}$, two $F=2$ atoms $\beta_{22}$, and collisions between atoms of different internal states $\beta_{12}$, as well as $p$ which is the pump rate of atoms transferred to the $F=2$ hyperfine state. The single particle loss was omitted in the rate equation since tests showed that fit values for the single particle loss coefficient did not differ from zero within the assumed uncertainties.

\begin{figure}
  \includegraphics[width=0.85\columnwidth]{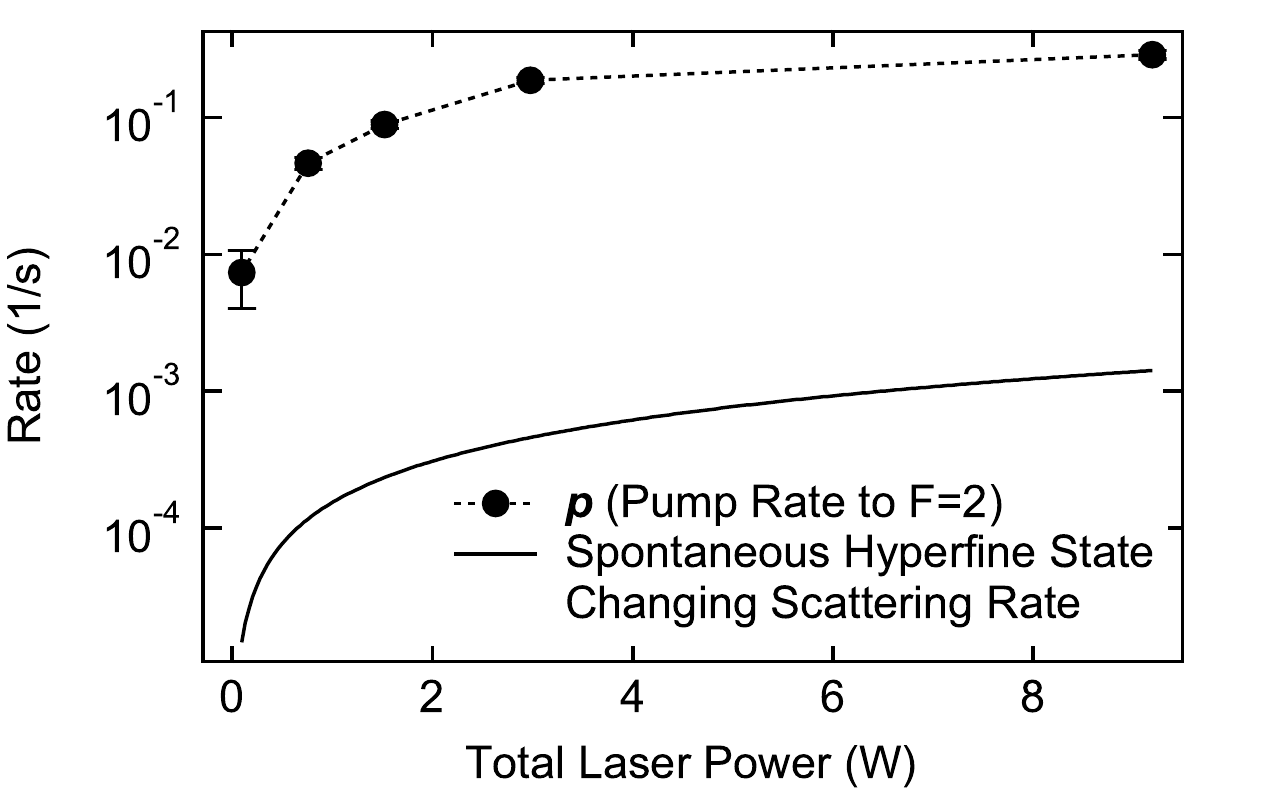}
\caption{Extracted pump rate $p$ and calculated rate of spontaneous Raman scattering events causing a change in the hyperfine state of an atom as function of trap laser power.}
\label{figpump}
\end{figure}

Figure \ref{figpump} presents the extracted pump rate $p$ together with the calculated rate of spontaneous Raman scattering events causing a change in the hyperfine state per atom. Since the latter rate is almost three orders of magnitude smaller than the pumping rate we encounter in our experiment, this again proofs that there have to be other than spontaneous Raman processes causing the rapid relaxation between the two hyperfine states. The observed dependence of the pump rate $p$ on the dipole trap laser power confirms that the light pumping the atoms to state $F=2$ is indeed the dipole trapping light and not other resonant stray light that might be introduced from other parts of the experiment. To ensure that there is no near-resonant light emitted by the fiber laser, we additionally spectrally filtered the fiber laser light: two mirrors, designed for high reflectivity between $780\,$nm and $830\,$nm and high transmittance at $1030\,$nm where added to the laser beam line. This results in an additional attenuation of at least a factor of $100$ for light at wavelengths between $762\,$nm and $905\,$nm and more than a factor of $5000$ attenuation for light resonant to the rubidium transitions at $780\,$nm and $795\,$nm, while the fiber laser light at $1070\,$nm is transmitted by $98\%$. The measured evolution of $N_1$ and $N_2$ showed no difference compared to the unfiltered case at the same power. Using all these observations, we conclude that the pumping to state $F=2$ is not caused by near resonant light accidentally emitted by the fiber laser or scattered from elements in the rest of the experiment.

\begin{figure}
  \includegraphics[width=0.7\columnwidth]{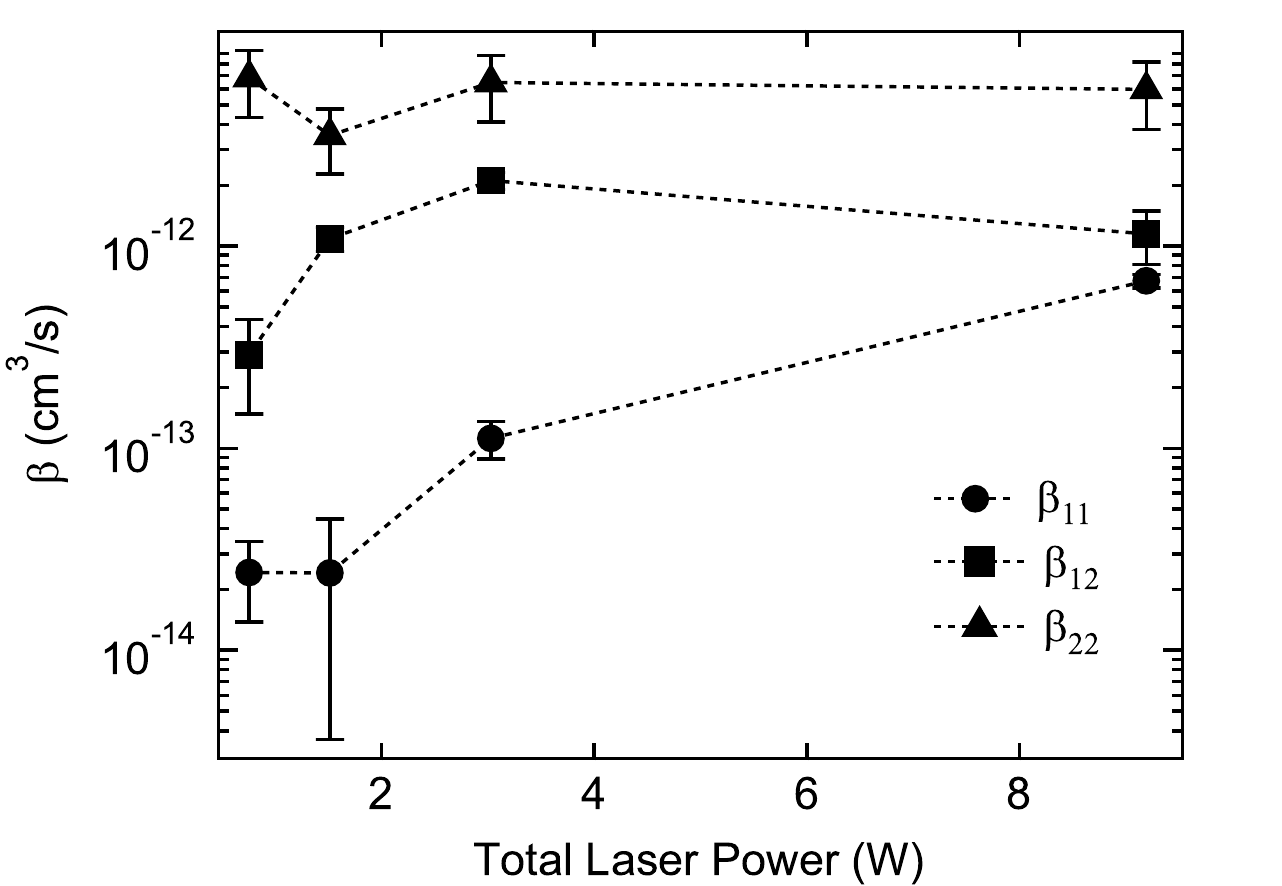}
\caption{Power dependent two-body loss coefficients $\beta$ extracted from atom loss measurements as shown in Fig. \ref{figf2pop} by fitting the solution of the rate equations (\ref{eqrate}). The error bars only represent statistical uncertainties. Expected systematic uncertainties are not included (see text).}
\label{figbeta}
\end{figure}

In Fig \ref{figbeta}, we present the results for the extracted $\beta$ coefficients as a function of laser power. The given uncertainties are statistical uncertainties of the fit only and do not include systematic uncertainties. The latter are significant, 
since the fit procedure shows a strong mutual dependence of the extracted values of the different $\beta$ coefficients. For that reason, we consider the given $\beta$ values only correct to within about one order of magnitude, and the following discussion relies in essence on the relative values of the coefficents and their general dependence on laser power.

The large ratio between the observed values of $\beta_{22}$ and $\beta_{12}$ on one side and $\beta_{11}$ on the other side confirms the adverse effect of pumping atoms to state $F=2$. The ratios $\beta_{22}$/$\beta_{11}$ of about 100 and $\beta_{12}$/$\beta_{11}$ of about 10 for most laser powers, show that the fraction of atoms in $F=2$ should not exceed a value of about $10\%$ to avoid a significant enhancement of atom losses.
Together with the results on atom pumping displayed in Fig. \ref{figf2frac}, this confirms that additional losses can be kept small when keeping the ramp time constant well below 2 seconds for laser powers above $1.5\,$W. This verifies the experimentally determined values of the ramp time contants shown in Fig. \ref{figtau}.

Finally, we also notice a dependence of the two-body loss rate coefficients $\beta_{11}$ and $\beta_{12}$ on laser power. This suggests that there are light induced Raman processes during collisions to which many intermediate molecular states are contributing \cite{Miller1993} and which cause changes in the kinetic energy by the amount of the ground state hyperfine energy splitting. As a consequence the respective $\beta$ coefficients are modified by the presence of the multifrequency laser light as well. 
A detailed investigation of this process would go beyond the scope of this paper.

%***************************************
\section{Summary and Conclusions}
%***************************************
We have presented an optimized strategy for production of Bose-Einstein condensates of $^{87}$Rb in a crossed dipole trap using light from a 1070\,nm multifrequency fiber laser at a repetition rate of 3 per minute. The dipole trap setup is very simple and it is not necessary to implement more sophisticated schemes like recompressible traps or precooling in a magnetic trap for an efficient BEC production.

We observe high two-body losses for a multifrequency laser at high laser power which rapidly decrease when reducing the power. The comparison of the atom number decay to the one observed with a single-frequency laser indicates that the reason for high losses lies in the broad frequency distribution of the laser light. This is verified by the observed large rate for pumping atoms to the upper hyperfine state for multifrequency laser light. The observed rate is by orders-of-magnitude larger than the one calculated for spontaneous Raman scattering processes.

Based on these results, we obtain a strategy for choosing the ramp time constants $\tau_{ramp}$ for the linear segments of evaporation to be a constant factor on the order of 6 smaller than the atom decay time $\tau$, obtained at the corresponding laser power. This should give reasonable parameters for the evaporation sequence even in the presence of large atom losses. Optimization of the evaporation can be achieved with variation of only one free parameter, namely the ratio $\tau/\tau_{ramp}$.

\begin{acknowledgements}
We wish to thank Jan Sch\"utz for insightful discussions. 
This work was supported financially in part by the RTN 'ATOM CHIPS' of the European Commission.
\end{acknowledgements}

\bibliographystyle{apsrev4-1}
\bibliography{dipolfibreBEC}
%\insert{dipolfibreBEC.bbl}
\end{document}